\documentclass[aps,prl,twocolumn,groupedaddress,showpacs]{revtex4}
\usepackage{epsfig}
\usepackage{graphicx}% Include figure files
\usepackage{dcolumn}% Align table columns on decimal point
\usepackage{bm}% bold math
\usepackage{appendix}

\newcommand{\la}{\langle}
\newcommand{\ra}{\rangle}

\newcommand{\vk}{{\bf k}}

\newcommand{\vq}{{\bf q}}
\renewcommand{\vr}{{\bf r}}

\begin{document}
\title{Spin-spectral-weight distribution and energy range of the parent compound La$_2$CuO$_4$}

\author{M. A. N. Ara\'ujo$^{1,2}$, J. M. P. Carmelo$^{3,4}$, M. J. Sampaio$^{3}$, and S. R. White$^{5}$}
\affiliation{$^{1}$Departamento de F\'{\i}sica, Universidade de \'Evora, P-7000-671, \'Evora, Portugal}
\affiliation{$^{2}$ CFIF, Instituto Superior T\'ecnico, Universidade T\'ecnica de Lisboa,  Av. Rovisco Pais, 1049-001 Lisboa, Portugal}
\affiliation{$^{3}$ Center of Physics, University of Minho, Campus Gualtar, P-4710-057 Braga, Portugal}
\affiliation{$^{4}$Institut f\"ur Theoretische Physik III, Universit\"at Stuttgart, D-70550 Stuttgart, Germany}
\affiliation{$^{5}$ Department of Physics and Astronomy, University of California, Irvine, CA 92617, USA}

\date{27 January 2012}

\begin{abstract}
The spectral-weight distribution in recent neutron scattering experiments on the parent compound La$_2$CuO$_4$ (LCO), 
which are limited in energy range to about 450\,meV, is studied in the framework of the Hubbard model on the square lattice. 
We find that the higher-energy weight extends to about 566\,meV and is located at and near the momentum $[\pi,\pi]$. 
Our results confirm that the $U/t$ value suitable to LCO is in the range $U/t\in (6,8)$. 
The continuum weight energy-integrated intensity vanishes or is extremely small at momentum $[\pi,0]$. This behavior of the 
intensity is consistent with that of spin waves, which are damped at $[\pi,0]$.
\end{abstract}


\maketitle

It is natural to expect that a greater understanding of the physics of the
high-$T_c$ superconductor undoped parent compounds, such as La$_2$CuO$_4$ (LCO), 
will lead to a greater understanding
of the corresponding superconductors. In particular, the less complicated undoped systems
can provide valuable information on which model Hamiltonians quantitatively describe
the cuprates. Improved determination of the model Hamiltonians is essential because of
the many nearby competing phases in the doped systems, easily affected by small parameters, 
which can now be seen because of 
continued improvements in numerical simulations \cite{WSstripe}.
A decade ago the neutron scattering experiments on LCO of Coldea, et. al. \cite{LCO-2001} first showed 
sufficient details of the spin-wave spectrum to demonstrate
that a simple nearest-neighbor Heisenberg model must be supplemented by a number of additional
terms, including ring exchanges. 
These terms arise naturally out of a single band Hubbard model with finite $U/t$,
and several detailed studies showed that the spin-wave data in the available energy window 
could be successfully described by the Hubbard model
using a somewhat smaller value of $U/t \sim 6-8$ than originally thought
appropriate \cite{LCO-2001,peres2002,lorenzana2005,companion}.
However, part of the spectral 
weight was deduced to be outside the energy window. 

Recently, improved neutron scattering experiments \cite{headings2010} with a much wider
energy window of about 450\,meV, have raised a number of questions. 
Surprisingly, these studies revealed that the high-energy spin waves are 
strongly damped near momentum $[\pi,0]$ and merge into a momentum-dependent continuum.
These results led the authors of Ref. \cite{headings2010} to conclude that
``the ground state of La$_2$CuO$_4$ contains additional correlations not captured by 
the N\'eel-SWT [spin-wave theory] picture''.
This raises the important question of whether the more detailed results can still be
described in terms of a simple Hubbard model. Here we address this question using a combination
of a number of theoretical and numerical approaches, including, in addition to standard
treatments, a new spinon approach for the spin excitations \cite{companion,companion0} and density matrix renormalization
group (DMRG) calculations for Hubbard cylinders \cite{Steve-92,DMRG-SL-2011,DMRG-tJ-2011}.
We show that the Hubbard model {\it does}  describe the new neutron scattering results.
In particular, at momentum $[\pi,0]$ the continuum weight energy-integrated 
intensity is found to vanish or be extremely small.
Furthermore, 
we find that beyond 450\,meV,  the spectral weight is mostly located around momentum $[\pi,\pi]$ 
and extends to about 566\,meV, suggesting directions for future experiments.
\begin{figure}[hbt]
\begin{center}
\centerline{\includegraphics[width=6.00cm]{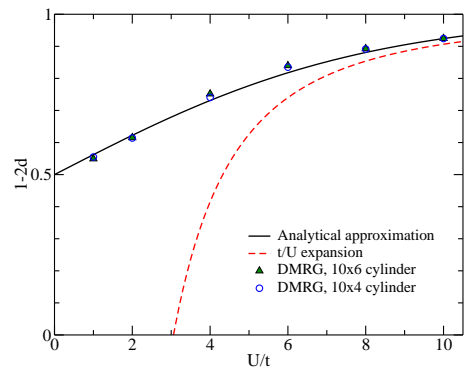}}
\caption{Average single-occupancy: approximate 
expression $[1+\tanh (U/8t)]/2$ valid for $U/t\leq 8$ (full line),
the limiting $U/t\gg 1$ expression $[1 -c_0 (8t/U)^2]$ (dashed line),
and from DMRG numerical results on two different width cylinders.}
\label{fig1}
\end{center}
\end{figure}

The Hubbard model on a square lattice with $N\gg 1$ sites and periodic boundary conditions reads $\hat{H} = \hat{T} + U\hat{D}$. Here
$\hat{T} = -t\,\sum_{\langle j,j'\rangle}\sum_{\sigma =\downarrow,\uparrow}[c_{\vr_j,\sigma}^{\dag}\,c_{\vr_{j'},\sigma}+h.c.]$
is the kinetic-energy operator, ${\hat{D}} = \sum_{j=1}^{N}\hat{n}_{\vr_j,\uparrow}\hat{n}_{\vr_j,\downarrow}$ 
counts the number of electron doubly occupied sites, $c_{\vr_j,\sigma}^{\dag}$ and $c_{\vr_{j},\sigma}$ are electron creation and 
annihilation operators with site index $j=1,...,N$ and spin $\sigma =\uparrow,\downarrow$, and
$\hat{n}_{{\vr}_j,\sigma} = c_{\vr_j,\sigma}^{\dag} c_{\vr_j,\sigma}$. The expectation values,
\begin{eqnarray}
d & = & \langle\hat{n}_{\vr_j,\uparrow}\hat{n}_{\vr_j,\downarrow}\rangle 
\, , \hspace{0.25cm}
(1-2d)  = \langle (\hat{n}_{\vr_j,\uparrow}-\hat{n}_{\vr_j,\downarrow})^2\rangle \, ,
\nonumber \\
m_{AF} & = & {1\over 2}\langle (-1)^j (\hat{n}_{\vr_j,\uparrow}-\hat{n}_{\vr_j,\downarrow})\rangle
\approx  [1-2\delta S]\,m_{AF}^0 \, ,
\label{d-m}
\end{eqnarray}
play an important role in our study, following the strong evidence that for $U>0$ and 
$N\rightarrow\infty$ the model ground state has antiferromagnetic long-range order \cite{Mano}. 
In the last expression of Eq. (\ref{d-m}), $m_{AF}^0$ stands for
a mean-field sublattice magnetization that does
not account for the effect of transverse fluctuations and $\delta S$
accounts for such an effect, its value being estimated below. Moreover,
${\hat{S}}^{z}_{\vr_j}={1\over 2}[{\hat{n}}_{\vr_j,\uparrow}-{\hat{n}}_{\vr_j,\downarrow}]$,
${\hat{S}}^{x}_{\vr_j}={1\over 2}[{\hat{S}}^{+}_{\vr_j} +{\hat{S}}^{-}_{\vr_j} ]$,
${\hat{S}}^{y}_{\vr_j}={1\over 2i}[{\hat{S}}^{+}_{\vr_j} -{\hat{S}}^{-}_{\vr_j} ]$,
${\hat{S}}^{+}_{\vr_j}=c_{\vr_j,\uparrow}^{\dag}\,
c_{\vr_j,\downarrow}$, and ${\hat{S}}^{-}_{\vr_j} = c_{\vr_j,\downarrow}^{\dag}\,c_{\vr_j,\uparrow}$.

Our study involves the dynamical structure factors,
\begin{eqnarray}
S^{\alpha\alpha'} (\vk,\omega) & = & {(g\mu_B)^2\over N}\sum_{j,j'=1}^{N}e^{-i\vk (\vr_j-\vr_{j'})}
\nonumber \\
& \times &
\int_{-\infty}^{\infty}dt\,e^{i\omega t}
\langle GS\vert {\hat{S}}^{\alpha}_{\vr_j} (t) {\hat{S}}^{\alpha'}_{\vr_{j'}} (0)\vert GS\rangle \, ,
\label{SS}
\end{eqnarray}
where $\alpha =\alpha'=x,y,z$ or $\alpha =-$ and $\alpha' =+$ and
below we consider $g=2$. It is straightforward to show that the sum rules
$[1/N]\sum_{\vk}\,S^{\alpha\alpha'} (\vk)$ 
where $S^{\alpha\alpha'} (\vk)= [1/2\pi]\int_{-\infty}^{\infty}d\omega\,S^{\alpha\alpha'} (\vk,\omega)$
involve the average single occupancy $(1-2d)$ and read
$[(g\mu_B)^2/4][\delta_{\alpha,\alpha'}+2\delta_{\alpha,-}\delta_{\alpha',+}] (1-2d)$.
In an ideal experiment one sees a transfer of weight from the longitudinal
to the transverse part such that independent of the scattering geometry, the corresponding
effective spin form factor satisfies the sum rule,
\begin{equation}
{1\over N}\sum_{\vk}{1\over 2\pi}\int_{-\infty}^{\infty}d\omega\,
S^{exp} (\vk,\omega) = \mu_B^2\,2(1-2d) \, .
\label{sr-eff}
\end{equation}
That the coefficient involved is $2(1-2d)$ rather than $3(1-2d)$ follows from in the experiment 
one mode being perpendicular to the plane and thus silent \cite{lorenzana2005}. 
\begin{figure}[hbt]
\begin{center}
\centerline{\includegraphics[width=6.00cm]{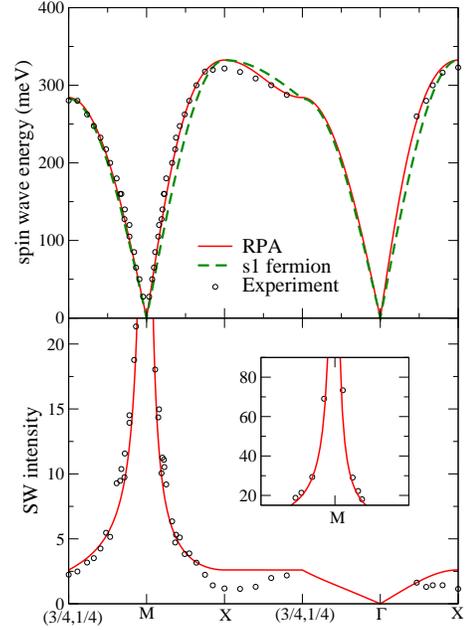}}
\caption{Upper panel: Spin-wave excitation spectrum along Brillouin-zone special directions  
as specified in Ref. \cite{headings2010}.
Lower panel: Spin-wave intensity as obtained from the poles of the susceptybility (see text).
Experimental points from Ref. \cite{headings2010}.}
\label{fig2}
\end{center}
\end{figure}

For $U/t\in (0,8)$ the antiferromagnetic long-range order may be accounted for by a variational 
ground state with a SDW initial trial state, such as 
$\vert G\rangle =e^{-g\hat{D}}\vert SDW\rangle$ or $\vert GB\rangle =e^{-h\hat{T}/t}e^{-g\hat{D}}\vert SDW\rangle$,
where $\vert SDW\rangle$ is the ground state of a simple effective mean-field Hamiltonian, as that of Eq. (18) of Ref. \cite{Dionys-09}.
For $U/t\gg 1$ such an order is as well accounted 
for by a Baeriswyl variational state $\vert B\rangle =e^{-h\hat{T}/t}\vert\infty\rangle$ where
$\vert\infty\rangle$ is the exact $U/t\rightarrow\infty$ ground state \cite{Dionys-09}.
The coefficients $h$ and $g$ multiplying the kinetic-energy and
double-occupancy operators, respectively, are variational parameters. 
Similarly for the trial state $\vert SDW\rangle$, the relation $4[m_{AF}^0]^2=(1-4d)$ holds for 
$\vert G\rangle$, $\vert GB\rangle$, and $\vert B\rangle$, while the function $d=d(U/t)$
is state dependent. This gives $d = {1\over 4}[1-4[m_{AF}^0]^2]$, 
consistent with $d$ not being affected by transverse fluctuations.

The evaluation of the ground-state energy for $\vert GB\rangle$ and $\vert B\rangle$ is for $N\gg 1$ 
an involved problem. Here we resort to an approximation, which corresponds to the simplest 
expression of the general form $E/N = T_0\,q_U  + U d$ where $T_0 = -{16\over\pi^2}\,t$
compatible with three requirements: the relation $d = {1\over 4}[1-4[m_{AF}^0]^2]$; the occurrence of
antiferromagnetic long-range order for the whole $U/t>0$ range; and the lack of a linear
kinetic-energy term in $U$ for $U/t\ll 1$. Brinkman and Rice found $q_U=8d (1-2d)$ 
for the original Gutzwiller approximation \cite{BR-70}, which is lattice insensitive and thus does not 
account for the square-lattice antiferromagnetic long-range order.  
The simplest modified form of the quantity $q_U$ such that the three conditions are met is 
$q_U = \left({U\over 8t}\right)\,a_{1}^{(+)}\,d\left[{(1-2d)\over  4[m_{AF}^0]^2}- a_2\right] - a_3$ 
where $4[m_{AF}^0]^2$ behaves as $4[m_{AF}^0]^2=U/8t$ for $U/t\ll 1$,
$a_{1}^{(\pm)} = \pi^2 \pm 4$, $a_2 = (1-[\pi^2/2a_1^{(+)}])$, and $a_3 = a_3 (U/8t)$
a function given by $a_3 = [a_{1}^{(-)}/8]\{1-\tanh ([U/8t][(4+a_{1}^{(+)})/a_{1}^{(-)}])\}$ for 
$U/t\in (0,8)$ and $a_3 =-c_0[\pi/2]^2\,[8t/U]$ for $U/t\gg 1$. Here $c_0 =[\alpha/4 +1/8]/2=0.1462$ and 
the corresponding estimate $\alpha =0.6696$ is that of the Heisenberg-model studies of Ref. \cite{Mano}.
Minimization of the obtained ground-state energy with respect
to $d$ leads indeed to $d = {1\over 4}[1-4[m_{AF}^0]^2]$.
For $U/t\ll 1$ such a $q_U$ expression yields to second order in
$U/t$, $E/N = T_0 + U d - [1/8\pi^2][U^2/t]$
where $[1/8\pi^2]\approx 0.0127$. The coefficient $\approx -0.0127$ is that also obtained 
by second-order perturbation theory \cite{MZ-89}. For $U/t\gg 1$ one recovers the known
result $E/N =-4c_0[8t^2/U]$ \cite{Mano}, so that our approximation agrees with
the known limiting behaviors.

For $U/t\in (0,8)$, we find that $4[m_{AF}^0]^2\approx \tanh (U/8t)$ gives quantitative agreement 
for the $(1-2d)$ dependence on $U/t$ with both our numerical DMRG calculations (see Fig. 1)
and
the numerical results for the states $\vert G\rangle$ and $\vert GB\rangle$
(see Fig. 4 of Ref. \cite{lorenzana2005}). In the DMRG calculations, 
two different circumference cylinders were simulated as a function $U/t$,
with open boundary conditions in $x$ and periodic in $y$, and the double occupancy measured in one
of the middle columns. A maximum of 
$m=4000$ states were kept, with an accuracy of $\sim 10^{-4}$ in $(1-2d)$ for the $10\times4$ system
for the least accurate smaller $U/t$ values,
and about $10^{-3}$ for the $10\times6$ system.
We find that the value of $(1-2d)$ is relatively insensitive to cluster size, and these cluster
sizes are representative of 2D behavior \cite{Paiva}.
For $U/t\gg 1$ we find the behavior
$4[m_{AF}^0]^2\approx e^{-2c_0\,(8t/U)^2}$ for the state $\vert B\rangle$, so that
$(1-2d) = [1+\tanh (U/8t)]/2$ for approximately $U/t \leq 8$ and $(1-2d) =[1 -c_0 (8t/U)^2]$
for $U/t\gg 1$.  Furthermore, the 
states $\vert SDW\rangle$ and $\vert B\rangle$ give $m_{AF}\approx m_{AF}^0={1\over 2}\sqrt{1-4d}$, with
an improved $d =d(U/t)$ dependence for the latter, whose $m_{AF}^0$ magnitudes are provided below
in Table \ref{table1} for several $U/t$ values. The 
states $\vert GB\rangle$ and $\vert B\rangle$ have $m_{AF}^{GB}$ and $m_{AF}^{B}$  
sublattice magnetization numerical values very close to those given by Eq. (\ref{d-m}) with $\delta S \approx d$ 
and $\delta S \approx d+{1\over 2}[1-m_{HAF}/m_{HAF}^0]\approx d+0.197$,
respectively. Here $m^0_{HAF}=1/2$ and $m_{HAF} \approx 0.303$ is the Heisenberg-model sublattice magnetization  
magnitude \cite{Mano}. Some $m_{AF}^{GB}$ magnitudes are given below in Table \ref{table1}, along with those
of $m_{AF}^{lower}=(1-2d)[m_{HAF}/m_{HAF}^0]\,m^0_{AF}$,
which we define for $U/t>0$ and for $U/t\gg 1$ becomes $m_{AF}^{B}$. Probably $m_{AF}^{lower}$ is
closest to the exact $m_{AF}$, while $m_{AF}^{GB}$ is consistent with our use of the RPA.
\begin{figure}[hbt]
\begin{center}
\centerline{\includegraphics[width=6.5cm]{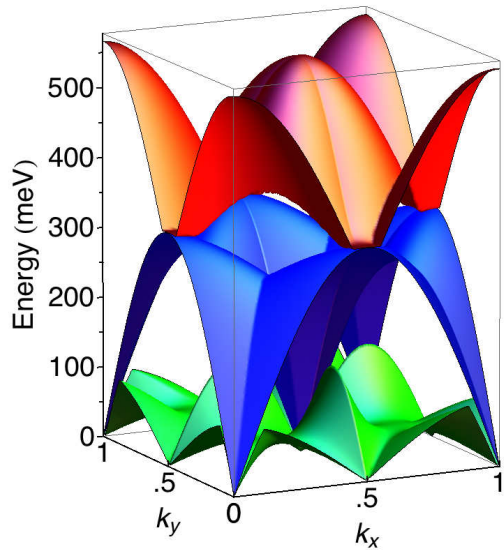}}
\caption{The energy-momentum space limits of the spin $S=1$ excited states spectrum for $U/t=6.1$, 
$t=295$\,meV, and $k_x$ and $k_y$ in units of $2\pi$. States whose energy is for a given $\vk$ lower than
that of the intermediate spin-wave sheet as well as those of any energy and equivalent momenta 
$[0,0]=[0,2\pi]=[2\pi,0]=[2\pi,2\pi]$ do not contribute to the spin spectral weight.}
\label{fig3}
\end{center}
\end{figure}
  
To study the coherent spin-wave weight distribution and spectrum, we have calculated by RPA
the transverse dynamical susceptibility
$\chi^{-+} (\vk,\tau)={(g\mu_B)^2\over N} \sum_{j,j'=1}^{N} e^{-i\vk\cdot\left(\vr_j-\vr_{j}'\right)} \la \hat S^-(\vr_j,\tau) \hat S^+(\vr_{j'},0)\ra$,
where $\tau$ denotes the imaginary time in Matsubara formalism (we shall take the zero temperature limit). 
In the case of the present model, the coherent spin-wave spectral weight in units of $\mu_B^2$ is
$Z_d\,2(1-2d)$. Here $Z_d = 1 - [2/N (1-2d)]\sum_{\vk}\sum_{\nu'\neq \nu}\vert\langle\nu'\vert {\hat{S}}^{+}_{\vk}\vert GS\rangle\vert^2$
where the sum over energy eigenstates excludes those  that generate the coherent spin-wave 
weight, $\vert\nu\rangle = \vert\nu,\omega (\vk)\rangle$. 
In the $U/t\rightarrow\infty$ limit, $Z_d$ may be identified with the corresponding
$Z_d = Z_c\,Z_{\chi}$ factor of the Heisenberg model on the square lattice. 
According to the results of Ref. \cite{Canali}, 
$Z_c\approx 1.18$ and  $Z_{\chi}\approx 0.48$, respectively, so that $Z_d \approx 0.57$. 
The limiting values $Z_d=1$ for $U/t\rightarrow 0$ and $Z_d \approx 0.57$ for $U/t\rightarrow\infty$
and the approximate intermediate value $Z_d\approx 0.65$ at $U/t=8$ \cite{lorenzana2005} are recovered
as solutions of the equation $Z_d = e^{-Z_d\tanh \left(\sqrt{U\over 4\pi t}\right)}$, which is used here
for finite $U/t$. In the thermodynamic limit the upper-Hubbard band processes generate nearly no spin weight.
Hence the longitudinal (elastic contribution to) spectral weight within $\mu_B^2\,2(1-2d)$, Eq. (\ref{sr-eff}), is 
in units of $\mu_B^2$ approximately given
by $\approx 4(m_{AF})^2$ and the spin-wave intensity reads $W_{SW} = Z_d\,[2(1-2d)-4(m_{AF})^2]$.
The GA+RPA method used in Ref. \cite{lorenzana2005} accounts for the quantum fluctuations that control 
the longitudinal and transverse relative weights, so that the spin-wave intensity factor is $Z_d$. The RPA 
used here leads to a similar spin-wave intensity momentum distribution but its experimentally determined factor
$Z_d^{exp}<Z_d$ is such that $W_{SW} = Z_d\,[2(1-2d)-4(m_{AF})^2]=Z_d^{exp}\,2(1-2d)$.

Within the two-sublattice description the susceptibility becomes a $2\times 2$ tensor and the
above original susceptibility $\chi^{-+}(\vk,\tau)$ is the average of its four elements.  
After Fourier transforming to $(\vk,i\omega)$ space, within the RPA the susceptibility tensor obeys
$\tilde\chi^{RPA} = [1-U \tilde\chi^{(0)}]^{-1} \tilde\chi^{(0)}$.
Here $\tilde\chi^{(0)}$ denotes the  susceptibility tensor in the noninteracting system. 
$\tilde\chi^{RPA} $ has a  pole $i\omega =\omega(\vk)$  obtained from the equation 
${\rm Det}\,[1-U \tilde\chi^{(0)}]=0$, which provides the dispersion relation $\omega(\vk)$ for the spin waves.
It has been shown in Ref. \cite{peres2002} that an excellent agreement with the spin-wave spectrum 
from Ref. \cite{LCO-2001} is achieved. In Fig. \ref{fig2} upper panel we show a fit to the 
more recent experimental data of Ref. \cite{headings2010} (full line) along with
the results from the $s1$ fermion method reported below (dotted line) for $U/t=6.1$ and $t=295$\,meV. Provided that the $t$ magnitude 
is slightly increased for increasing values of $U/t$, agreement with the LCO spin-weight spectrum and distribution can be 
obtained for $U/t\in (6,8)$. For $U/t$ values smaller than $6$ (and larger than $8$), the spin-wave 
dispersion between $[\pi/2,0]$ and $[\pi/2,\pi/2]$ has a too large energy bandwidth (and is too flat) for any 
reasonable value of $t$. From the residue of the spin-wave pole the susceptibility coherent part reads,
\begin{equation}
\chi_{co}^{-+} (\vk,i\omega) = Z_d^{exp}\sum_{l=\pm1}
\frac{{\rm Res}\,[\chi^{-+}(\vk,l\,\omega(\vk)] }{i\omega-l\,\omega(\vk)} \, .
\label{completechi}
\end{equation}
The measured intensity is $I_{SW}(\vk)=\pi [S^{xx}(\vk)+S^{yy}(\vk)]=\pi S^{-+}(\vk)$ \cite{authors}.
In Fig. \ref{fig2} lower panel, we plot the corresponding RPA
spin-wave intensity, $I_{SW}(\vk)=-[\pi/2] Z_d^{exp}{\rm Res}\,[\chi^{-+} (\vk,\omega(\vk)]$. The good agreement 
with the experimental data, specially near the point $M$, reproduces the theoretical results of Ref. \cite{headings2010}.
It is here obtained for the value $Z_d^{exp}\approx 0.49$,
which corresponds to the choice $m_{AF}=m_{AF}^{GB}= m_{AF}^{0}(1-2d)$ such that
$\delta S\approx d$. As in Ref. \cite{headings2010}, it shows disagreement around the $X$ point, which here 
probably stems from the RPA.

To derive the spectrum of the spin $S=1$ excited states producing
the inelastic form-factor spectral weight, we use the spinon
operator representation of Ref. \cite{companion}. The ground-state spin degrees of freedom are
described by a full $s1$ fermion band, which coincides with an antiferromagnetic reduced Brillouin 
zone. Each $s1$ fermion is a spin-singlet two-spinon composite object. The 
spin $S=1$ excited states of momentum $\vk=[\pi,\pi] -\vq-\vq\,'$
are generated upon breaking one $s1$ fermion spinon pair, which leads to the emergence
of two holes of momenta $\vq$ and $\vq\,'$ in the otherwise full $s1$ band. The inelastic coherent spin-wave
spectrum is generated by processes where $\vq$ points in the nodal direction and $\vq\,'$
belongs to the boundary of the $s1$ band reduced zone \cite{companion}. The remaining
choices of $\vq$ and $\vq\,'$ either generate the inelastic incoherent continuum spectral weight
or vanishing weight, respectively. The studies of Ref. \cite{companion} are limited to the spin-wave spectrum. Here we consider the
energy-momentum space domain of all above spin $S=1$ excited states, which is represented in Fig. \ref{fig3}.
(A similar spectrum is obtained for the values $U/t=8.0$ and $t=335$\,meV 
of Ref. \cite{lorenzana2005}.) The intermediate sheet refers to the spin-wave spectrum. 
For each $\vk$, states of energy lower than the latter spectrum do not contribute 
to the form-factor weigh. Furthermore and consistent with the first-moment sum rules of an isotropic antiferromagnet, 
no and nearly no weight is generated by states of any energy and momentum $[0,0]$ and near $[0,0]$, respectively.
Unfortunately, the method of Ref. \cite{companion} does not provide
the detailed continuum weight distribution. However, it is expected that, similarly to the Heisenberg model case \cite{lorenzana2005}, 
its energy-integrated intensity follows the same trend as the spin-wave intensity. 
Analysis of Fig. \ref{fig3} reveals that for momentum $[\pi,0]$ there are no excited states of energy higher than
the spin waves. Thus at momentum $[\pi,0]$, the continuum weight distribution energy-integrated 
intensity vanishes or is extremely small, due to $s1$ band four-hole processes. Given the expected common trend of both intensities,
this is consistent with a damping of the spin-wave intensity at momentum $[\pi,0]$, as observed \cite{headings2010}
but not captured by the Fig. \ref{fig2} RPA intensity. 
\begin{table}
\begin{tabular}{|c|c|c|c|c|c|} 
\hline
$U/t$ & $6.1$ & $6.5$ & $8.0$ & $10.0$ \\
\hline
$W_T$ & $1.643$ & $1.671$ & $1.762$ ($1.778$ \cite{lorenzana2005}) & $1.848$ ($1.846$ \cite{lorenzana2005}) \\
\hline
$W_{SW}$ & $0.808$ & $0.799$ & $0.761$ & $0.714$ \\
\hline
$W_{<450}$ & $1.571$ & $1.593$ & $1.663$ & $1.730$ \\
\hline
$W_{>450}$ & $0.072$ & $0.078$ & $0.099$ & $0.118$ \\
\hline
$m^{0}_{AF}$ & $0.401$ & $0.410$ & $0.436$ ($0.43$ \cite{lorenzana2005,Dionys-09}) & $0.461$ ($0.456$ \cite{lorenzana2005}) \\
\hline
$m^{GB}_{AF}$ & $0.329$ & $0.342$ & $0.384$ ($0.39$ \cite{Dionys-09}) & $0.426$ \\
\hline
$m^{lower}_{AF}$ & $0.200$ & $0.207$ & $0.233$ & $0.258$ \\
\hline
\end{tabular}
\caption{Several spectral weights in units of $\mu_B^2$ as defined in the text and the
sublattice magnetizations as calculated here for several $U/t$ values and
some results from Refs. \cite{lorenzana2005,Dionys-09}.}
\label{table1}
\end{table} 

In Table \ref{table1} we provide the sublattice magnetization magnitudes along 
with our calculations of the following integrated spectral weights (in units of $\mu_B^2$):
the total weight, $W_T=2(1-2d)$; the spin-wave weight, 
$W_{SW} = Z_d^{exp}\,2(1-2d)$; the weight
for energy $\hbar\omega\leq$450 meV, $W_{<450}=W_{SW}/0.71 +4(m_{AF}^{GB})^2$; and
the weight $W_{>450}=[W_T-W_{<450}]$ for energy $\hbar\omega >$ 450 meV.
$W_{<450}$ is derived by combining our theoretical expressions with the 
observations of Ref. \cite{headings2010} that for the energy range up to about 450 meV,
71\% and 29\% of the weight corresponding to the inelastic response comes from the
coherent spin-wave weight and incoherent continuum weight, respectively.
Our prediction for $W_{<450}$ varies between $1.6\,\mu_B^2$ for $U/t\approx 6.1$ and $1.7\,\mu_B^2$ for $U/t\approx 8.0$,
in agreement with the experimental value $1.9\pm 0.3\,\mu_B^2$ reported in
Ref. \cite{headings2010}. From our above analysis,
the small weight $W_{>450}\approx 0.1\,\mu_B^2$ is expected to extend to about 566 meV
(see Fig. \ref{fig3}), mostly at and around the momentum $[\pi,\pi]$.

Our result that at momentum $[\pi,0]$ the continuum weight energy-integrated intensity vanishes or is very small
is consistent with a corresponding damping of the spin-wave intensity at $[\pi,0]$, as observed 
in the recent experiments of Ref. \cite{headings2010}. We suggest that future LCO neutron 
scattering experiments scan the energies between 450 meV and 
566 meV and momenta around $[\pi,\pi]$.

We thank the authors of Ref. \cite{headings2010} for
providing their experimental data and A. Muramatsu for discussions. 
J. M. P. C. thanks the hospitality of the University of Stuttgart
and the support of the German Transregional Collaborative 
Research Center SFB/TRR21 and Max Planck Institute for Solid State Research. 
S.R.W. acknowledges the support of the NSF under DMR 090-7500.

\end{document}